\pdfoutput=1
\documentclass[conference]{IEEEtran}
\IEEEoverridecommandlockouts
\usepackage{cite}
\usepackage{amsmath,amssymb,amsfonts}
\usepackage{algorithmic}
\usepackage{graphicx}
\usepackage{textcomp}
\usepackage{xcolor}
\usepackage{amsmath}
\def\BibTeX{{\rm B\kern-.05em{\sc i\kern-.025em b}\kern-.08em
    T\kern-.1667em\lower.7ex\hbox{E}\kern-.125emX}}
\usepackage[nolist]{acronym}
\usepackage{subfig}
\usepackage{tikz}
\usepackage{bm}
\usepackage{booktabs} 
\usepackage{tabularx}
\begin{document}

\onecolumn
\section*{IEEE COPYRIGHT NOTICE}
\textcopyright 2020 IEEE. Personal use of this material is permitted. Permission from IEEE must be obtained for all other uses, in any current or future media, including reprinting/republishing this material for advertising or promotional purposes, creating new collective  works,  for  resale  or  redistribution  to  servers  or  lists,  or  reuse  of  any  copyrighted  component  of  this  work  in  other works.

\textbf{Accepted to be published in: Proceedings of the 2020 IEEE International Symposium on Hardware Oriented Security and Trust  (HOST),  December  6-9,  2020,  San  Jose,  CA,  USA}
\twocolumn

\title{Encoding Power Traces as Images for Efficient Side-Channel Analysis}

\author{\IEEEauthorblockN{Benjamin Hettwer\IEEEauthorrefmark{1}\IEEEauthorrefmark{2}, Tobias Horn \IEEEauthorrefmark{3}, Stefan Gehrer\IEEEauthorrefmark{4}, Tim G\"uneysu\IEEEauthorrefmark{2}}
	\IEEEauthorblockA{\IEEEauthorrefmark{1}Robert Bosch GmbH, Corporate Sector Research, Renningen, Germany}
	\IEEEauthorblockA{\IEEEauthorrefmark{2}Horst G\"ortz Institute for IT-Security, Ruhr University Bochum, Germany}
	\IEEEauthorblockA{\IEEEauthorrefmark{3}Esslingen University of Applied Sciences, Esslingen, Germany}
	\IEEEauthorblockA{\IEEEauthorrefmark{4}Robert Bosch LLC, Corporate Sector Research, Pittsburgh, USA}	
		\\ Email: benjamin.hettwer@de.bosch.com, tohoit06@hs-esslingen.de, stefan.gehrer@bosch.com,
		\\ tim.gueneysu@rub.de} 

\maketitle

\begin{abstract}
\acp{SCA} are a powerful method to attack implementations of cryptographic algorithms. State-of-the-art techniques such as template attacks and stochastic models usually require a lot of manual preprocessing and feature extraction by the attacker. \ac{DL} methods have been introduced to simplify \acp{SCA} and simultaneously lowering the amount of required side-channel traces for a successful attack. However, the general success of \ac{DL} is largely driven by their capability to classify images, a field in which they easily outperform humans. 

In this paper, we present a novel technique to interpret 1D traces as 2D images. We show and compare several techniques to transform power traces into images, and apply these on different implementations of the \ac{AES}. By allowing the neural network to interpret the trace as an image, we are able to significantly reduce the number of required attack traces for a correct key guess. We also demonstrate that the attack efficiency can be improved by using multiple 2D images in the depth channel as an input. Furthermore, by applying image-based data augmentation, we show how the number of profiling traces is reduced by a factor of 50 while simultaneously enhancing the attack performance.
This is a crucial improvement, as the amount of traces that can be recorded by an attacker is often very limited in real-life applications.

\end{abstract}

\begin{IEEEkeywords}
Side-channel attacks (SCAs), Deep Learning (DL), AES, Convolutional Neural Network (CNNs).
\end{IEEEkeywords}

\hyphenation{hard-ware}

\begin{acronym}
	\acro{SCA}[SCA]{Side-Channel Attack}
	\acro{EM}[EM]{Electromagnetic}
	\acro{TA}[TA]{Template Attack}
	\acro{ML}[ML]{Machine Learning}
	\acro{DL}[DL]{Deep Learning}
	\acro{DNN}[DNN]{Deep Neural Network}
	\acro{CNN}[CNN]{Convolutional Neural Network}
	\acro{PDF}[PDF]{Probability Density Function}
	\acro{POI}[POI]{Points of Interest}
	\acro{MLP}[MLP]{Multi-Layer Perceptron}
	\acro{ReLU}[ReLU]{Rectified Linear Unit}
	\acro{SGD}[SGD]{Stochastic Gradient Descent}
	\acro{CONV}[CONV]{Convolutional}
	\acro{POOL}[POOL]{Pooling}
	\acro{FC}[FC]{Fully-Connected}
	\acro{SOFT}[SOFT]{Softmax}
	\acro{IN}[IN]{Input}
	\acro{ACT}[ACT]{Activation}
	\acro{HW}{Hamming Weight}
	\acro{OUT}[OUT]{Output}
	\acro{KGE}[KGE]{Key Guessing Entropy}
	\acro{GPU}{Graphical Processing Unit}
	\acro{AES}{Advanced Encryption Standard}
	\acro{SNR}{Signal-to-Noise Ratio}
	\acro{DK}{Domain Knowledge}
	\acro{LL}{Log-likelihood}
	\acro{KRPC}{Key Rank Perturbation Curve}
	\acro{ZB-KGE}{Zero-Baseline Key Guessing Entropy}
	\acro{POIs}{Points of Interest}
	\acro{SOST}{Sum Of Squared pairwise T-differences}
	\acro{SOSD}{Sum Of Squared Differences}
	\acro{PCA}{Principal Component Analysis}
	\acro{TVLA}{Test Vector Leakage Assessment}
	\acro{CPA}{Correlation Power Analysis}
	\acro{DOM}{Difference of Means}
	\acro{MIA}{Mutual Information Analysis}
	\acro{ILSVRC}{ImageNet Large Scale Visual Recognition Challenge}
	\acro{GAF}{Gramian Angular Field}
	\acro{GASF}{Gramian Angular Summation Field}
	\acro{GADF}{Gramian Angular Difference Field}
	\acro{MTF}{Markov Transition Field}
	\acro{RP}{Recurrence Plot}
	\acro{LDA}{Linear Discriminant Analysis}
	\acro{STFT}{Short Time Fourier Transform}
	\acro{DFT}{Discrete Fourier Transform}
	\acro{PAA}{Piecewise Aggregate Approximation}
	\acro{SSA}{Singular Spectrum Analysis}
	\acro{FFT}{Fast Fourier Transform}
\end{acronym}

\section{Introduction}
Since the advent of \acfp{SCA} by Kocher et al. in 1996 \cite{Kocher_1996}, the topic has become a serious threat for security devices and is extensively investigated in research and industry. \acp{SCA} take advantage of information leakages through timing, power consumption or \ac{EM} emanations to extract secret parameters from cryptographic algorithms implemented in software or hardware. Numerous attack methods have been proposed in literature to analyze side-channel leakages. These can be divided into two major categories: Non-profiled and profiled \acp{SCA}. Techniques such as \ac{CPA} \cite{Brier2004} or \ac{MIA} \cite{Gierlichs_08} belong to the non-profiled class and aim to recover cryptographic keys by using statistical calculations between real power measurements and a hypothesis of the leakage behaviour of the device under attack. In profiled \acp{SCA}, the adversary performs an additional profiling step with an open copy of the target device ahead of the actual attack using Gaussian templates \cite{Chari2003}, stochastic attacks \cite{Schindler2005}, or machine learning methods \cite{Hettwer_19}.

Another path of work dealing with \acf{DL}-based profiled side-channel analysis has been introduced by Maghrebi et al. in 2016 \cite{Maghrebi2016}. Several publications showed that especially \acp{CNN} are able to outperform classic key recovery techniques (see, e.g., \cite{Cagli2017} \cite{Hettwer_18} \cite{Kim_2019} for an incomplete list).
\acp{CNN} are special types of \ac{DL} models that were originally proposed for image recognition tasks \cite{Le_cun_98}. They exploit local connection patterns in the input data through receptive fields and weight sharing, and are - to some degree - shift, scale, and distortion invariant. 
State-of-the-art \ac{CNN} architectures for image classification like VGG have been winning the \ac{ILSVRC} since 2012, partly showing above human-level performance.
In the context of \acp{SCA}, \acp{CNN} were mainly applied on the raw, one-dimensional side-channel traces. However, there are several ways to transform 1D data structures (like side-channel traces) to 2D images in order to enable visual learning techniques to recognize patterns in time series data. Hatami et al., e.g., proposed \acp{RP} to transform time series into 2D texture images \cite{Hatami_17}. Audio or sound classification can be performed with spectrograms in order to identify features in the time-frequency domain \cite{COSTA_2017}. Wang and Oates introduced \acp{GAF} and \acp{MTF} to encode time series as images and achieved state-of-the art results on 20 different data sets \cite{Wang_15}.

\begin{figure*}
	\centering
	\includegraphics[scale=0.75, bb=0 0 500 150]{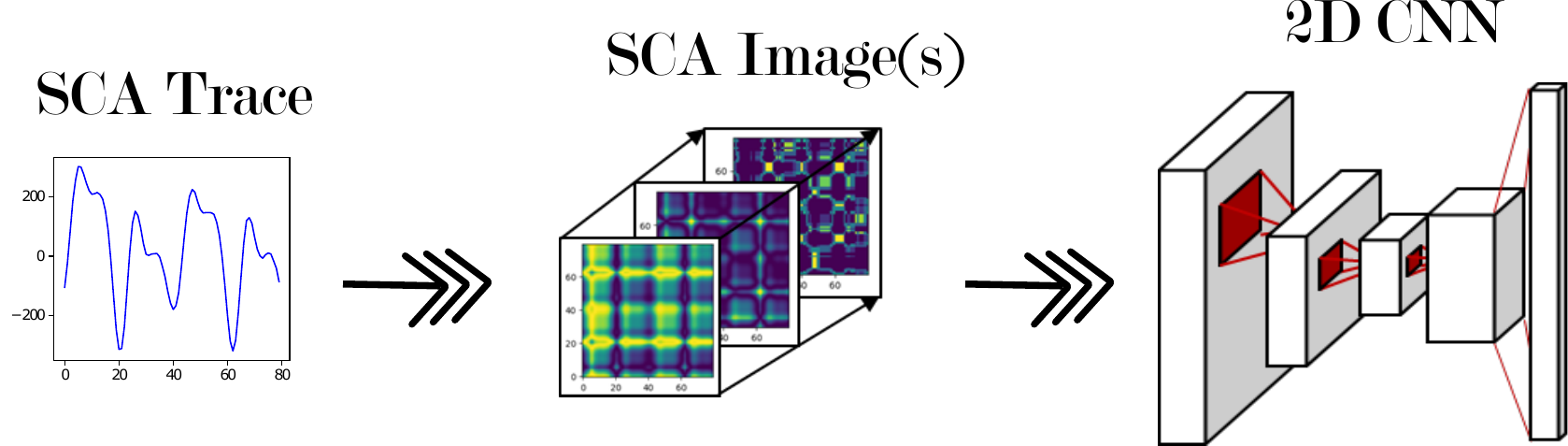}
	\caption{Top-level overview of our approach. One-dimensional \ac{SCA} traces are transformed into images, which are later used as input for a 2D \ac{CNN}. It is also possible to stack several pictures generated from the same trace like different channels in an RGB image.}
	\label{fig:overview}
\end{figure*}

Motivated by the success of \acp{CNN} in computer vision, this paper investigates the performance of profiled \acp{SCA} in 2D space. We encode power traces from four different data sets as images using \ac{GAF}, \ac{MTF}, \acp{RP} and spectrograms, and apply them to a 2D \ac{CNN} in order to extract secret key bytes from cryptographic implementations. While there are notable differences between the methods, our experiments show that 2D-based attacks can be more powerful than their equivalents in 1D space. This holds also in case the traces are preprocessed with state-of-the-art 1D transformations such as \ac{SSA} \cite{Merino_15}. We furthermore demonstrate how the performance of the \ac{CNN} is further enhanced by combining different 2D representations of a trace into a single image as illustrated in \figurename{~\ref{fig:overview}}. Finally, we explore image-based data augmentation to boost the robustness of our \ac{CNN} model and report valid parameter settings for all transformation techniques.

The remaining structure of the paper is as follows: In Section \ref{sec:rel_work}, we give an overview about related work. In Section \ref{sec:prelim}, we shortly recap profiled \acp{SCA} and introduce 2D \acp{CNN}. In Section \ref{sec:image}, we present several techniques to transform time series to images and apply them to power traces. In Section \ref{sec:exp}, we evaluate our approach by attacking unprotected and protected \ac{AES} implementations. The last section summarizes the paper and gives insights on possible future work.

\section{Related Work}
\label{sec:rel_work}
There have been numerous tools and techniques proposed in literature that deal with the preprocessing of side-channel traces to increase the power of \acp{SCA}. They can be grouped in the following categories:

\subsection{Denoising}
Techniques to improve the \ac{SNR} range from simple trace averaging to advanced filtering using Kalman and Wiener filters \cite{Souisse_10}, or the forth-order cumulant to remove Gaussian noise \cite{Le_07}. Noise reduction in the frequency domain is also widely used in practice, e.g., wavelet-based methods \cite{Liu_14}, Hilbert transform \cite{Yao_16} and \ac{SSA} \cite{Merino_15}. Maghrebi et al. recently proposed Independent Component Analysis, a technique for blind source separation, to denoise side-channel measurements and could show a significant \ac{SNR} gain \cite{Maghrebi_18}. The methods investigated in this paper are not explicitly designed to decrease the level of noise in time series, but can be easily combined with them before the transformation in the 2D space is done.

\subsection{Dimensionality Reduction and POI Selection}
Identifying \ac{POIs} or features in side-channel traces has been traditionally studied in the context of \acp{TA}, in order to reduce the computational overhead during calculation of the covariance matrices. In particular, \ac{DOM} \cite{Chari2003}, \ac{SOST}, \ac{SOSD} \cite{gierlichs_06}, and \ac{PCA} \cite{Archambeau_06} have been proposed for that purpose. Another common strategy for \ac{POIs} selection and dimensionality reduction is based on Pearson correlation, whereby the importance of sample points is measured by the correlation coefficient of the actual power consumption and some key-dependent target intermediate value \cite{Mangard2010}. 

\subsection{Dataset Augmentation}
\acp{DNN} usually require a large amount of training data in order to learn effectively. Since generating more data is often too expensive or not possible at all, fake data can be created and added to the training set which is referred to as dataset augmentation \cite{Goodfellow_2016}. It has been particularly effective in object recognition tasks, where images are, e.g., flipped, rotated or cropped to boost the performance of the classifier \cite{Taylor_17}.
Cagli et al. investigated data augmentation in the context of \acp{SCA} to defeat jitter-based countermeasures by artificially distorting traces  \cite{Cagli2017}. To this end, shifting deformations (simulates random delay) and add-remove deformations (simulates clock jitter) were applied to the traces of different training sets. Pu et al. furthermore introduced a lazy validation method to find an appropriate range of which the traces can be shifted horizontally \cite{Pu_17}. Recently, Kim et al. proposed to add Gaussian noise to the input during training in order to improve the attack performance \cite{Kim_2019}. 
In this work, we apply several data augmentation techniques to 2D representations of side-channel traces and compare them with the aforementioned methods from the 1D domain.
\section{Preliminaries}
\label{sec:prelim}

\subsection{Profiled Side-Channel Attacks}
\label{sec:sca}
Profiled attacks are considered as the most powerful type of \acp{SCA} and are composed of two phases: a \textit{profiling} (or training) phase, and an \textit{attack} phase (inference in machine learning context). During the first phase, the attacker uses a profiling device (which is similar to the target device and on which input parameters as the secret key and plaintext can be set arbitrarily) to build a very precise model of the physical leakage. For that purpose, he acquires a set of $S_P$ side channel traces  $\mathbf{x} \in \mathbb{R}^N$ from the profiling device, where $N$ denotes the number of sample points in the measurements. Let $y = g(p,k)$ be a random variable representing the result of an intermediate operation of the target cipher which depends partly on public information $p$ (plaintext or ciphertext chunk) and secret key (byte) $k \in \mathcal{K}$, where $\mathcal{K}$ is the set of possible key values. The value of $y$ is assumed to have an influence on the deterministic part of the power or \ac{EM} measurements. As a final step, the adversary estimates the probability distribution:
\begin{equation}
	\text{Pr}[\mathbf{x}|Y=y]
\end{equation}
using the profiling set $ \mathcal{D}_{Profiling} = \{(\mathbf{x}_i, y_i)\}_{i=1}^{S_P}$.

In the attack phase, a new set $\mathcal{D}_{Attack}$ composed of $S_A$ attack traces from the actual target device is acquired by the adversary. The target device is structurally identical to the profiling device whereby the secret key $k$ is fixed but unknown. In order to recover $k$, the adversary follows a maximum likelihood strategy and estimates the posterior probability for all possible key candidates $k^*$ using Bayes' theorem:
\begin{equation}
	\mathbf{d} = \prod_{i=1}^{S_A} \text{Pr}[Y=y_i|\mathbf{x}_i] 
	 = \prod_{i=i}^{S_A} \frac{\text{Pr}[\mathbf{x}=\mathbf{x_i}|y_i=g(p_i,k^*)]}{\text{Pr}[y_i=g(p_i,k^*)]}
	 \label{eq:max_lh}
\end{equation}

where the $k$-th entry in the score vector $\mathbf{d}$ corresponds to the correct key candidate \cite{Prouff_18}.

\subsection{\acfp{CNN}}
\label{sec:cnn}
\acp{CNN} incorporate dedicated knowledge about a specific type of input into their architecture. They are designed to process data that has a grid like topology such as time-series data or images. Such data usually has a strong local structure meaning that variables (or pixels) that are spatially or temporally nearby are highly correlated \cite{Le_cun_98}. Unlike traditional learning algorithms, \acp{CNN} are able to automatically extract and combine local and global features. The main building blocks for that purpose are the \textit{convolution} and \textit{pooling} layers which are introduced in the following subsections. Afterwards, we give a brief overview about the principal architecture and training of \acp{CNN}. We refer the reader to \cite{Goodfellow_2016} for a comprehensive description of \acp{CNN} and \ac{DL} in general.

\subsubsection{Convolutional Layer}
This type of layer is the core building block of a \ac{CNN} to exploit spatially local correlations. To this end, a kernel or filter $\mathbf{W}$ of size $r \times r$ (sometimes called receptive field) glides throughout the input $\mathbf{F}$ of size $m \times m$ and calculates an activation or feature map $\tilde{\mathbf{F}}$ according to the formula:
\begin{equation}
	\tilde{\mathbf{F}}(x,y) = \sum_{i=1}^{r} \sum_{j=1}^{r} \mathbf{F}(x+i,y+j)\cdot \mathbf{W}(i,j), 1 \leq x,y \leq m
	\label{eq:conv}
\end{equation}
There are usually several kernels in a convolutional layer, whereby every kernel creates a corresponding feature map which are stacked together along the depth dimension. The weight parameters of the kernels are learned to activate when they detect a specific feature or pattern in the input.

\subsubsection{Pooling Layer}
The pooling layers are responsible to reduce the resolution of the input in order to decrease the
number of parameters and the computational complexity of the network. In most \acp{CNN}, this is done by the max-pooling operation which considers only the maximum value within a rectangular neighborhood. Apart from downsampling the data stream, pooling helps the network to be invariant to small translations of the input, i.e., the exact location of a structure is not important but if the structure is present in the input at all.

\subsubsection{Architecture \& Learning}
Typical \ac{CNN} constructions consist of repetitive blocks of convolution and pooling layers. The basic concepts of sparse, local connectivity, weight sharing and subsampling enable the network to extract more abstract representations of given inputs. On top of the feature extractor part, fully connected layers, which are composed of neurons as in traditional neural networks, are responsible for feature combination and finally classification of the input. Non-linear activation functions, such as \textit{sigmoid} or \textit{\ac{ReLU}}, ensure that the network is able to approximate more complex dependencies. These are directly connected to the output of convolutional or fully-connected layers. The neurons in the last layer output class scores in form of a probability distribution by means of the Softmax function.

\acp{CNN} are usually trained in an iterative, multi-step process by which the weights of the kernels and neurons are optimized to minimize a dedicated loss function (i.e., the categorical cross-entropy in a classification setting). The loss depicts the difference between the expected output (i.e. labels) and the prediction result produced by the last layer. For faster convergence, gradient-based optimizer algorithms such as \textit{\ac{SGD}} or \textit{ADAM} are applied in practice for updating the parameters of the model. An important hyperparameter that controls the training process is the learning rate, which determines how fast the weights of the network are driven towards the optimal solution.

\section{Side-channel Analysis in 2D}
\label{sec:image}

In this section we introduce our method to create 2D representations of power traces. To that end, we investigate different signal processing techniques and give suitable parameter settings. Afterwards, we compare the methods with respect to the amount of leakage that can be detected in the generated images.

\subsection{Imaging Side-Channel Traces}

\newcolumntype{Y}{>{\centering\arraybackslash}X}

\begin{table*}[t]
	\centering
	\caption{Overview of transformation techniques into 2D space}
	\label{tab:images}
	\begin{tabularx}{\linewidth}{lYYYY}
		\toprule
		 & \textbf{\ac{GAF}} & \textbf{\ac{MTF}} & \textbf{\ac{RP}} & \textbf{STFT} \\
		 \midrule
		 \textbf{Encoded information} & Superposition/Difference of temporal correlations & Transition probabilities between quantiles & Phase space dynamics & Frequency per time \\
		 \textbf{Parameters} [\textit{used value}] & PAA [false] & Number of quantiles $Q$ [8]& Dimension of trajectories $M$ [1], delay $\tau$ [1], threshold $\epsilon$ [0], binarization [false]  & Window type [Hann],  window length $\triangle t$ [8], overlap [0.9$\cdot\triangle t$], sample frequency $F_S$ \mbox{[$1.25/2 \cdot 10^9$ Hz]} \\
		 \textbf{Image size} (for trace of length $N$) & $N \times N$ & $N \times N$ & $N \times N$ & $N \times 5$ \\
		\bottomrule
	\end{tabularx}
\end{table*}

In the following, we discuss four techniques to transform a vector $\mathbf{x} = [x_1, \dots, x_N] \in \mathbb{R}^N$ consisting of side-channel measurements into a matrix $\mathbf{X} \in \mathbb{R}^{M \times N}$, where $M$ and $N$ denote the height and width of the matrix. An overview of all techniques is provided in Table \ref{tab:images}. The motivation to encode \ac{SCA} traces as images is to generate new and more discriminative features that are not directly present in the time domain. This enables us to treat the problem of classifying power traces in a profiled \ac{SCA} as an image recognition task, which can be effectively handled by 2D \acp{CNN}. It is furthermore possible to concatenate different 2D representations of a single power trace into a multi-channel image to enhance the robustness and performance of the classifier.

\subsubsection{\acf{GAF}}
\ac{GAF} exploits the temporal correlation of the time series data to create images \cite{Wang_15}. For that purpose, a time series trace $\mathbf{x}$ is rescaled to a range of $[-1,1]$ using a Min-Max scaler:
\begin{equation} \label{eq:MinMax}
\tilde{x_i}= \frac{x_i - max(\mathbf{x}) + (x_i - min(\mathbf{x}))}{max(\mathbf{x}) - min(\mathbf{x})}
\end{equation}

Afterwards, the trace is transformed into the polar coordinate system:
\begin{equation} \label{eq:polar}
\begin{cases}
\phi = \arccos(\tilde{x_i}), -1\leq \tilde{x_i} \leq 1, \tilde{x_i}\in \tilde{\mathbf{x}}\\
r = \frac{t_i}{N}, t_i \in \mathbb{N}
\end{cases}
\end{equation}

The normalization of the trace in (\ref{eq:MinMax}) and the $\arccos$ in (\ref{eq:polar}) result in a bijective mapping in the polar coordinate system. The radius for the calculated value $\tilde{x_i}$ is determined through its point in time $t_i$ in the trace. This is important, because two different traces must not lead to the same image. Thus, the temporal correlations for each measurement are preserved. The last step is to calculate the Gram-Matrix $\mathbf{G}$ using the transformed values:
\begin{equation}\label{eq:GASF}
\mathbf{G} = \begin{bmatrix}
\cos(\phi_1 + \phi_1) &  \dots & \cos(\phi_1 + \phi_N) \\
\cos(\phi_2 + \phi_1)  & \dots & \cos(\phi_2 + \phi_N) \\
\vdots & \ddots & \vdots \\
\cos(\phi_N + \phi_1) & \dots & \cos(\phi_N + \phi_N) \\
\end{bmatrix}
\end{equation}
The usage of angles to create the Gram-Matrix leads to a non-Gaussian distribution. This can be shown, when the trigonometric sum, as defined in (\ref{eq:GASF}), is converted to the Cartesian coordinate system:
\begin{equation}
\begin{split}\label{eq:NonGaus}
\cos(\phi_1 + \phi_2) &= \cos(\arccos(x_1) + \arccos(x_2)) \\
&= \cos(\arccos(x_1)) \cdot \cos(\arccos(x_2)) \\
&	- \sin(\arccos(x_1)) \cdot \sin(\arccos(x_2)) \\
&= x_1 \cdot x_2 + \sqrt{1-x_1^2} \cdot \sqrt{1-x_2^2}
\end{split}
\end{equation}
As shown in (\ref{eq:NonGaus}), a penalty term $\sqrt{1-x_1^2} \cdot \sqrt{1-x_2^2}$ is added in the Cartesian system. It reaches its maximum when $x_1=x_2 = 0$. Another advantage of equation (\ref{eq:GASF}) is that the original time series value can be reconstructed through the main diagonal of the matrix.

The \ac{GAF} comes in two different versions. On the one hand as shown in equation (\ref{eq:GASF}), which is known as \ac{GASF} and on the other hand the \ac{GADF} which is defined as follows:
\begin{equation}\label{eq:GADF}
\mathbf{G} = \begin{bmatrix}
\sin(\phi_1 - \phi_1)&  \dots & \sin(\phi_1 - \phi_N) \\
\sin(\phi_2 - \phi_1)& \dots & \sin(\phi_2 - \phi_N) \\
\vdots & \ddots &  \vdots \\
\sin(\phi_N - \phi_1)& \dots & \sin(\phi_N - \phi_N) \\
\end{bmatrix}
\end{equation}


In contrast to \ac{GASF}, \ac{GADF} uses the difference of the values over the time $t$ to calculate the temporal correlation. Moreover, it is not possible to reconstruct the original trace through the main diagonal since:
\begin{equation}
 \sin(\phi_i-\phi_i) = 0, i = 1 \dots N
\end{equation}

The advantage of the transformation to \ac{GAF} images is that the temporal dependencies are preserved through superposition or differentiation. Additionally, there is no Gaussian distribution anymore which is important for \acp{CNN} to ease distinguishing relevant data from noise. However, a disadvantage of those transformations is the increased data overhead. A trace with $N$ sample points results in a $N \times\ N$ \ac{GAF} image. In \cite{Wang_15}, \ac{PAA} has been used to reduce the number of data points in time series. We have not applied this step since the averaging induced by \ac{PAA} would lead to undesired information loss. Instead, we decided to use only parts of the traces for the transformation (details will be explained later in the paper).
Examples of \ac{GAF} images are illustrated in \figurename{~\ref{fig:images}}b and \figurename{~\ref{fig:images}}c (first row).

\newcommand{\column}[2]{%
	\begin{tabular}[b]{@{}c@{}}#1\\#2\end{tabular}%
}

\begin{figure*}[h!]
	\centering
	\captionsetup[subfigure]{labelformat=empty}
	\begin{tabular}{cccccc}
	\column{
	\subfloat[]{\scalebox{.35}{\includegraphics[bb=0 0 200 150]{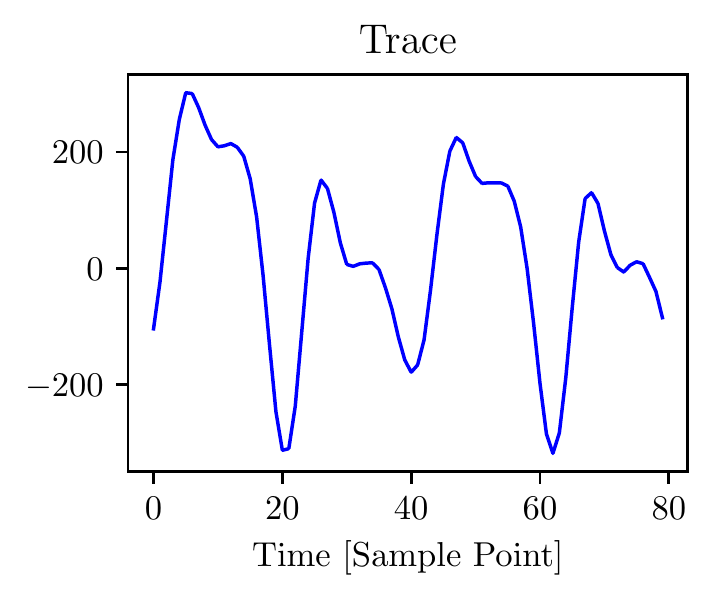}}\label{fig:trace}}\\
	\subfloat[]{\scalebox{.35}{\includegraphics[bb=0 0 200 175]{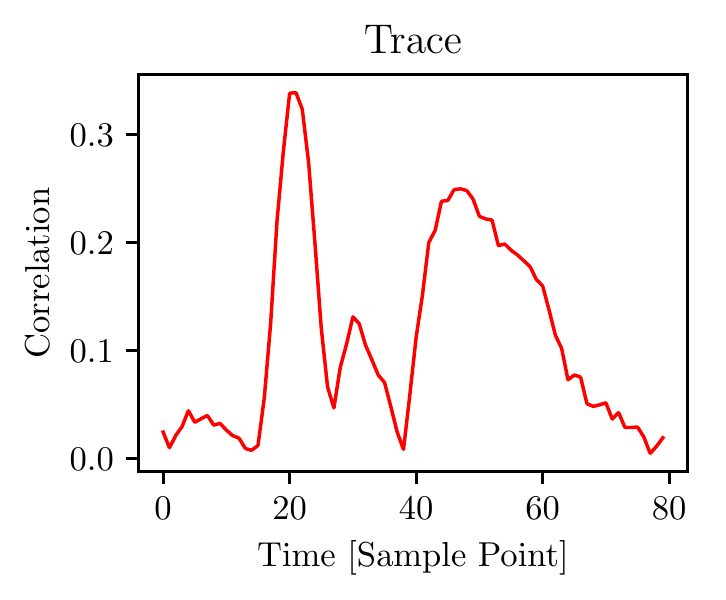}}\label{fig:corr_trace}}
	}{(a)} &
	\column{
	\subfloat[]{\scalebox{.35}{\includegraphics[bb=0 0 200 150]{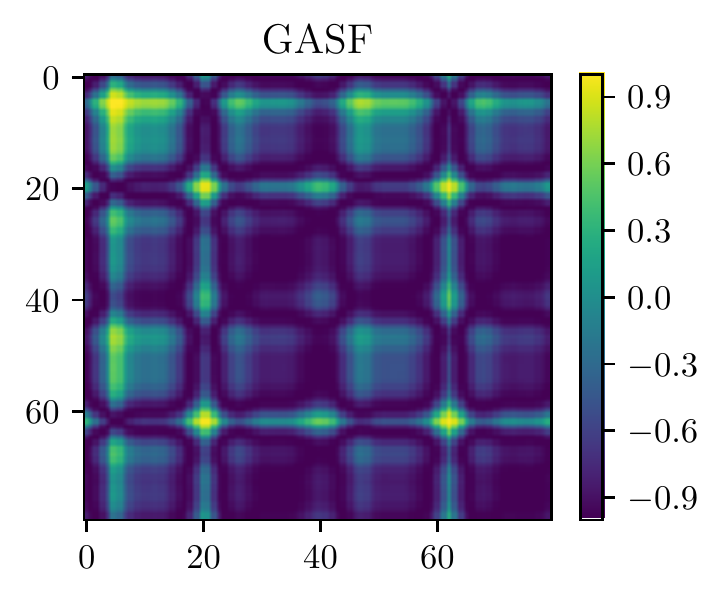}}\label{fig:gasf}} \\
	\subfloat[]{\scalebox{.35}{\includegraphics[bb=0 0 200 175]{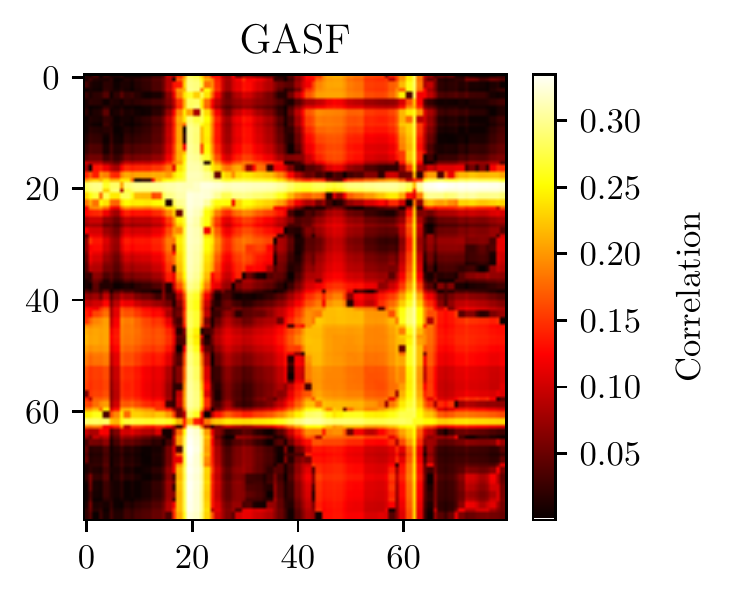}}\label{fig:corr_gasf}}
	}{(b)} &
	\column{
	\subfloat[]{\scalebox{.35}{\includegraphics[bb=0 0 200 150]{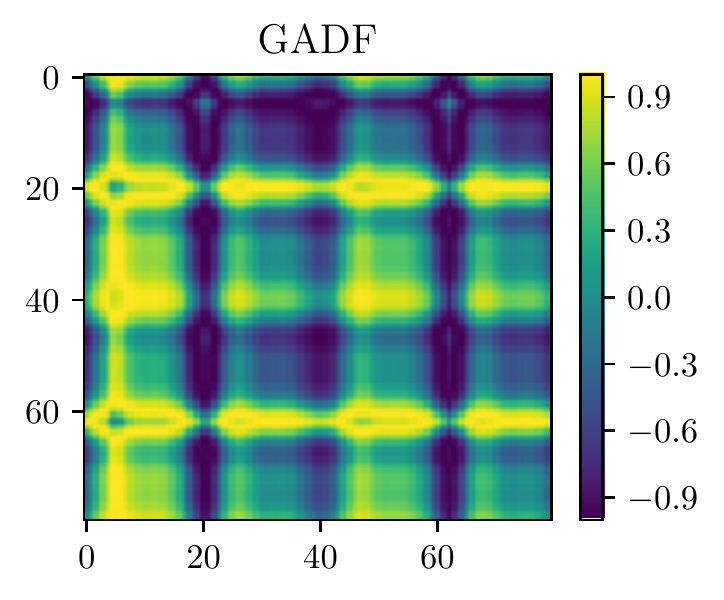}}\label{fig:gadf}} \\
	\subfloat[]{\scalebox{.35}{\includegraphics[bb=0 0 200 175]{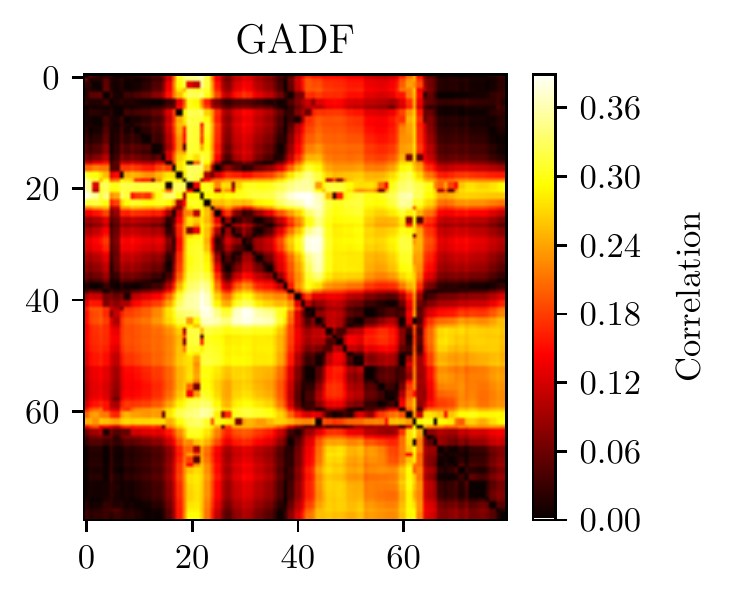}}\label{fig:corr_gadf}} 
	}{(c)} &
	\column{
	\subfloat[]{\scalebox{.35}{\includegraphics[bb=0 0 200 150]{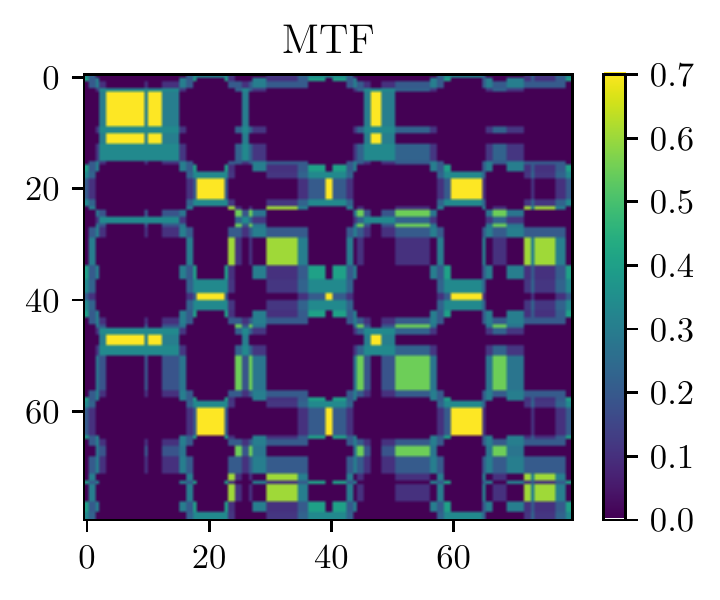}}\label{fig:mtf}} \\
	\subfloat[]{\scalebox{.35}{\includegraphics[bb=0 0 200 175]{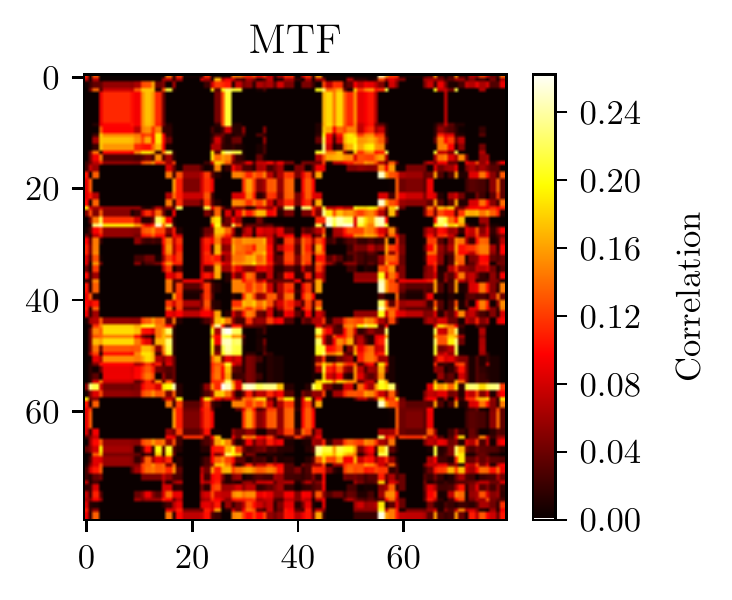}}\label{fig:corr_mtf}} 
	}{(d)} &
	\column{
	\subfloat[]{\scalebox{.35}{\includegraphics[bb=0 0 200 150]{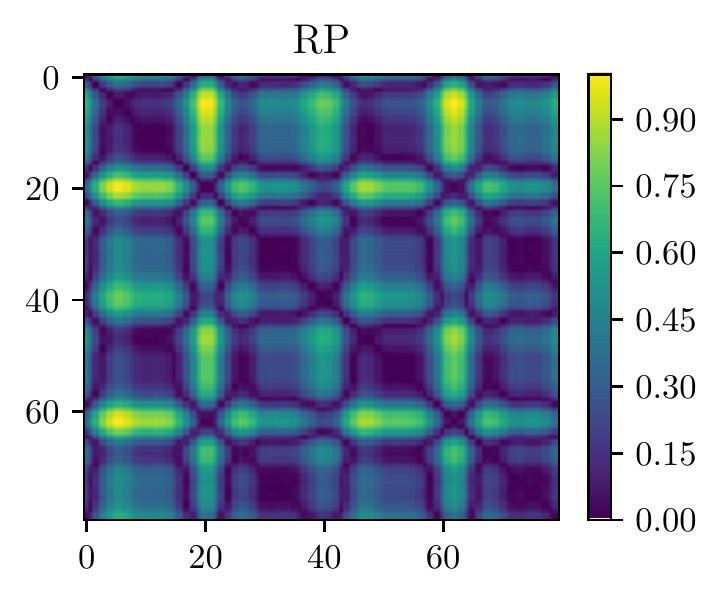}}\label{fig:rp}} \\
	\subfloat[]{\scalebox{.35}{\includegraphics[bb=0 0 200 175]{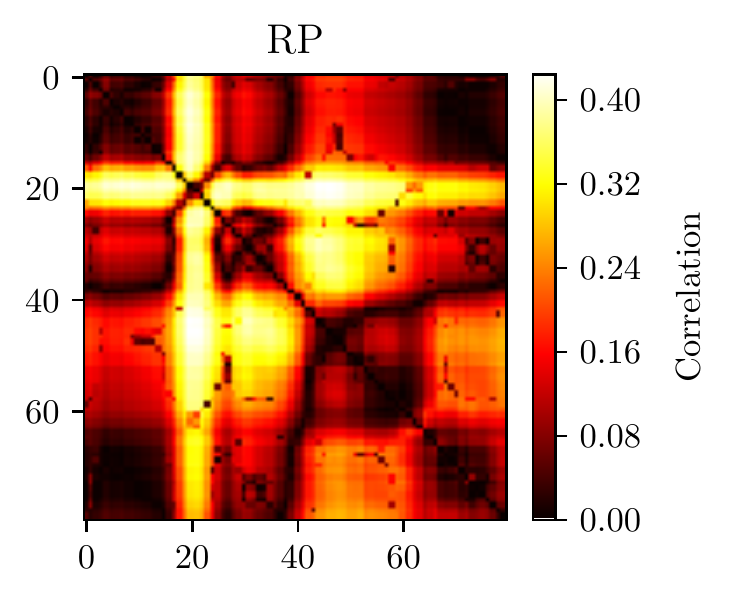}}\label{fig:corr_rp}}
	}{(e)} &
	\column{
	\subfloat[]{\scalebox{.35}{\includegraphics[bb=0 0 200 150]{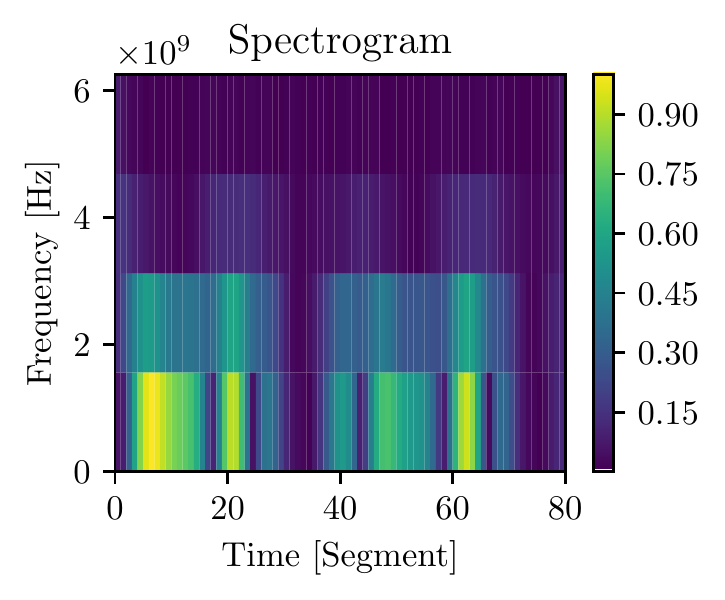}}\label{fig:spectrogram}} \\
	\subfloat[]{\scalebox{.35}{\includegraphics[bb=0 0 200 175]{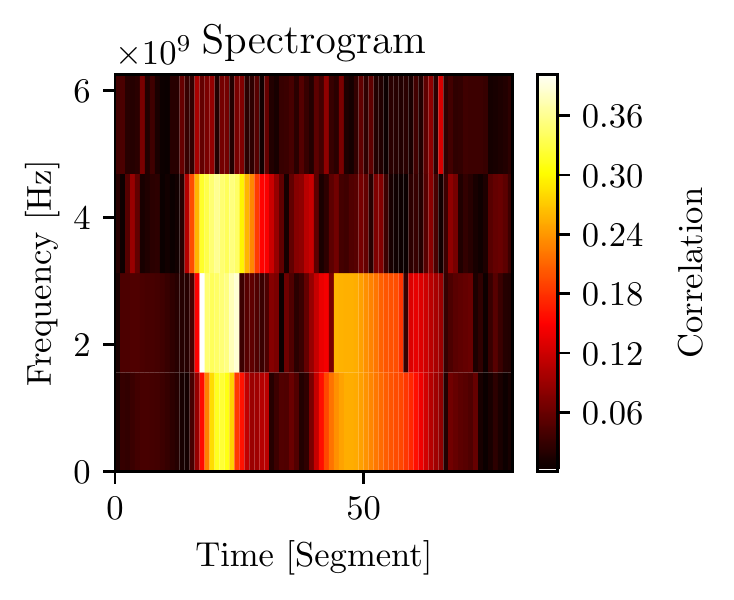}}\label{fig:corr_spectrogram}}
	}{(f)} 
	\end{tabular}

	\caption{Images created from power trace the AES-Serial data set: (a) original trace segment, (b) GASF , (c) GADF, (d) MTF, (e) RP, (f) spectrogram. The RP image and the spectrogram have been rescaled into a range between 0 and 1 manually after the transformation. The corresponding correlation plots are shown in the second row.}
	\label{fig:images}
\end{figure*}

\subsubsection{\acf{MTF}}
This is a another time series imaging method introduced by Wang et al. \cite{Wang_15}. In order to apply it for \acp{SCA}, the power trace $\mathbf{x}$ is divided into $Q$ quantile bins. Every sample point of $x_i$ is assigned to a bin $q_j (j \in [1,Q])$, which segments the data over its amplitude.
Subsequently a $Q \times Q$ weighted adjacency matrix $\mathbf{W}$ is created, by counting the transitions between the bins over time using a first-order Markov chain.
$w_{i,j}$ represents the frequency by which a point in quantile $q_j$ is followed by a point in quantile $q_i$. Through the normalization $\sum_{j} w_{ij} = 1$, the \ac{MTF} is created. It has the disadvantage that information about the temporal dependencies are lost. Hence, Wang et al. had the idea to define the \ac{MTF} as follows:
\begin{equation}\label{eq:MTF}
\mathbf{M} = \begin{bmatrix}
w_{ij|x_1\in q_i,x_1\in q_j} & \dots & w_{ij|x_1\in q_i,x_N\in q_j} \\
w_{ij|x_2\in q_i,x_1\in q_j} & \dots & w_{ij|x_2\in q_i,x_N\in q_j} \\
\vdots &  \ddots & \vdots \\
w_{ij|x_N\in q_i,x_1\in q_j} & \dots & w_{ij|x_N\in q_i,x_N\in q_j} \\
\end{bmatrix}
\end{equation}  
The quantiles which contain the values $x_i$ and $x_j$ at the time $i$ and $j$ are $q_i$ and $q_j$ ($q \in [1,Q]$). Thus, the \ac{MTF} $M_{ij}$ represents the transition probability $q_i \rightarrow q_j$. As a result, the \ac{MTF} shows for each point of time in the trace the probability of the transition from the quantile of the value $x_n \in q_i$ to the quantile of the value $x_m \in q_j$. Moreover, $M_{i,j||i-j|=l}$ represents  the transition probability within the time interval $l$. However, the main diagonal $M_{ii}$ is a special case which shows the probability of a transition to the same quantile with $l=0$. As a last step, to increase the efficiency of computation and to reduce the image size, an average filter with blurring kernel is applied on non-overlapping rectangles. The result of a \ac{MTF} transformation is visualized in \figurename{~\ref{fig:images}}d.

A crucial parameter in the context of \ac{SCA} is the number of quantile bins $Q$ used during the quantization. We investigated different values in the range $[5, N]$. Intuitively, a larger value leads to a finer subdivision of the signal which seems to be beneficial to capture small differences in the amplitude. However, we got the best results using eight quantiles.

\subsubsection{\acfp{RP}}
\acp{RP} have already been introduced by Eckmann et al. in 1987 \cite{Eckmann_1987}. The main idea is to visualize the M-dimensional phase space trajectory of a non-stationary time series signal as a two-dimensional representation of its recurrences. Informally spoking, an \ac{RP} indicates at which time instances the trajectory of the signal passes almost the same area in the phase space. It can be mathematically described as:

\begin{equation}
	\mathbf{R}_{i,j} = \mathcal{H}(\epsilon - \|\vec{x}_i - \vec{x}_j\|), \vec{x} \in \mathbb{R}^T, i,j = 1 \dots N_T
	\label{eq:rp}
\end{equation}

where $N_T$ is the number of considered trajectories $\vec{x}$, $\epsilon$ a threshold distance, $\|.\|$ the (Euclidean) norm, and $\mathcal{H}$ the Heaviside function. A single trajectory is defined by:

\begin{equation}
	\vec{x}_i = (x_i, \dots, x_{i+(M-1)\tau}), i = 1, \dots, N-(M-1)\tau
\end{equation}
where $M$ is the dimension of the trajectories and $\tau$ is the time gap between two back-to-back points of the trajectory.
\ac{RP} constructs a symmetric matrix that relates the distance between states as intensity values (i.e. the smaller the distance, the higher the intensity in the image).
Typical high-level patterns that can be found in \acp{RP} are referred to as topology and can be separated in the categories \textit{homogeneous, periodic, drift} and \textit{disrupted}. Due to the oscillating signature of power traces, \acp{RP} created from them contain large, recurrent structures as shown in the example in \figurename{~\ref{fig:images}}e. We have used a phase space dimension $M=1$ and time delay $\tau = 1$ throughout the experiments in order to capture fine-granular phase change information in the \acp{RP}. The choice of these parameter values is also supported by \ac{RP} literature \cite{Iwanski_98} and furthermore simplifies the combination with other imaging techniques, since the dimensionality of the resulting \ac{RP} is the same as for \ac{GASF}/\ac{GADF}. Additionally, we skipped the binarization induced by the Heaviside function in equation (\ref{eq:rp}) to avoid information loss.

\subsubsection{\acf{STFT}}
The Fourier transform decomposes a signal into its frequency components. This is shown by the \ac{DFT} in equation (\ref{eq:FFT}),  where $\mathbf{x}$ is the discrete and $F(\omega)$ the frequency transformed signal.

\begin{equation}\label{eq:FFT}
\begin{split}
F(\omega) &= \sum_{n=0}^{N-1}\mathbf{x}[n]e^{-i\frac{2\pi \omega n}{N}}, \omega=0, \dots, N-1
\end{split}
\end{equation} 

However, by transforming a signal into the frequency domain, time resolution is lost. Thus, it is known that a certain frequency occurred but not when in time. This poses a problem for \acp{SCA}, where it is beneficial to know within which time a certain frequency occurs. Therefore, as shown in equation (\ref{eq:STFT}), the \ac{STFT} does not transform the whole signal at once. Instead, a window function $w[n-m]$ is used which slides over the signal $\mathbf{x}[n]$. Only the selected area of the window is transformed into frequency domain using \ac{DFT}. By doing so, the frequencies within a time period of $\triangle t$ are received.
\begin{equation}\label{eq:STFT}
STFT(m,\omega) \equiv F(m,\omega) = \sum_{n=-\infty}^{\infty}\mathbf{x}[n]w[n-m]e^{-i\omega n}
\end{equation}

The spectrogram is a visualization of \ac{STFT} and plots the intensity of the \ac{STFT} magnitude over time.

STFT underlies the K\"{u}pfm\"{u}ller's uncertainty principle, which describes a correlation between the resolution of the time and the frequency domain. As a consequence, a high time resolution leads to a blurry resolution in the frequency domain and vice versa. That is because frequencies whose period length is larger than the window size $\triangle t$ cannot be resolved. In the context of \acp{SCA}, Yang et al. suggested to choose window lengths in a range between 64 and 256 to obtain a reasonable frequency resolution and balanced spectrogram image sizes \cite{Yang_18}. However, as we will demonstrate in the next section, also smaller window sizes which lead to spectrograms having the same time resolution than the original signal can give promising results. They have just to be adjusted to the special properties of \acp{CNN}. An example spectrogram using a window length of 8 is shown in \figurename{~\ref{fig:images}}f.

\subsection{Power Leakage Analysis in Images}
In order to compare the amount of information that is encoded by the different techniques, we have calculated the Pearson correlation between the generated images and leakage sensitive variable $y$. For that purpose, we first unrolled the images along the height dimension into 1D arrays. Once the correlation coefficient has been determined for individual pixels, we constructed a 2D matrix of the same size as the images.

The second row of \figurename{~\ref{fig:images}} illustrates the result using 1000 images per technique. 
One can observe that equivalent structures can be found in the images and corresponding correlations. Bright areas in the correlation plots indicate locations with higher leakage, and areas with low information content are depicted in black. It can be seen that \ac{GASF}, \ac{GADF} and \ac{RP} images show leakage in similar regions in the upper left part, while the leakage in \ac{MTF} images is more distributed. The leakage in spectrograms is most concentrated due to the small frequency resolution. Another interesting point can be noticed when comparing the highest amplitude of the correlation for the 1D traces (\figurename{~\ref{fig:images}}a) with the maximum correlations that can be found in the images. \acp{GADF} $(\approx 0.38)$, \acp{RP} $(\approx 0.42)$ and spectrograms $(\approx 0.38)$ seem to contain more information than the original traces $(\approx 0.33)$. However, \ac{MTF} shows a lower correlation maximum $(\approx 0.26)$ . The leakage of \ac{GASF} images is on a similar level as in the traces. From that result, we expect better results for at least some of the imaging techniques compared to \acp{SCA} in 1D.

\section{Experiments}
\label{sec:exp}
In this section, we investigate the performance of the aforementioned transformation techniques in the context of \ac{DL}-based \acp{SCA}.

\subsection{CNN Architecture}
The \ac{CNN} architecture we have used throughout all experiments is given in Table \ref{tab:cnn}. It consists of two blocks each of which contains a convolution and max-pooling layer followed by two fully-connected layers. Dropout layers and L2-regularization are included for better generalization of the model and help to reduce overfitting \cite{Goodfellow_2016}. Additionally, batch normalization is applied to establish a stable distribution of activation values, which makes the network more robust to parameter tuning. We have trained the \ac{CNN} using the Adam optimizer and a learning rate of 0.0002. In order to avoid overfitting, the weights leading to the smallest validation loss has been saved during training, and the process has automatically been stopped once the validation loss has not decreased for 20 consecutive epochs.

Our elaborated network configuration is inspired partly by \textit{VGGNet} \cite{VGG} which showed remarkable performance in image recognition. However, we are using fewer convolutional blocks and the fully-connected layers are of smaller size since our network only has to discriminate between $2^8 = 256$ different categories (classification of one byte) instead of 1000. We also experimented with a pre-trained feature extractor (convolution and pooling layers) of VGGNet, but got inconsistent results. This is potentially because VGGNet has been trained with real life objects, while our setup requires to classify time-series images which are composed of completely different structures and features. Please note that the goal of our paper is not to describe an optimal network architecture but rather to illustrates the advantages of \ac{SCA} in 2D domain in general.
Nevertheless, our proposed \ac{CNN} model is able to reach a high performance on all data sets.

\begin{table}
	\centering
	\caption{Network configuration of \ac{CNN}}
	\label{tab:cnn}
	\begin{tabular}{cc}
		\toprule
		\textbf{Layer Type} & \textbf{Hyperparameters}\\
		\midrule
		Convolution 2D & filters=8, kernel size=5x5, L2-reg=0.01\\
		Batch Normalization & - \\
		Activation & ReLU \\
		Max-Pooling & pool size=2x2, stride=2 \\
		Dropout & rate=0.5 \\
		Convolution 2D & filters=16, kernel size=3x3,  L2-reg=0.01 \\
		Batch Normalization & - \\
		Activation & ReLU \\
		Max-Pooling & pool size=2x2, stride=2 \\
		Dropout & rate=0.5 \\
		Flatten & - \\
		Fully-Connected & neurons=250, activation=ReLU\\
		Batch Normalization & - \\
		Dropout & rate=0.3 \\
		Fully-Connected & neurons=256, activation=Softmax\\
		\bottomrule
	\end{tabular}
\end{table}

\subsection{Data Sets}
We consider four different data sets: an unprotected hardware \ac{AES}, an AES hardware implementation with random jitter, a hardware AES setup with high noise, and a side-channel protected software \ac{AES}.
The size of the profiling and attack sets have been set to $S_P = 50\,000$ and $S_A = 10\,000$ examples for all data sets. We describe them briefly in the following.

\subsubsection{Unprotected Hardware AES}
The power traces of this data set originate from an unprotected hardware implementation of the \ac{AES} running on a ZYNQ UltraScale+ FPGA. The implementation contains a single S-Box instance, which is why we refer to this data set as AES-Serial. It is commonly known that the most leakage in a hardware implementation is caused by register transitions \cite{Mangard2010}. Therefore, we have used the XOR of two consecutive S-Box outputs in the first round as target operation and consequently as labels for training. A single measurement originally contains 1000 sample points, but we have only used a segment of 80 data points from the traces to generate the images (the location of these points was roughly determined by correlation with known key). Please note that the reduction has only been done in order to speed-up the experiments, we expect similar results using the complete traces.

\subsubsection{Hardware AES with desynchronization}
This data set is similar to the former one except that the traces have been artificially desynchronized. To this end, we have shifted each trace to the right with a random offset in the range of $[0, 50]$. Such misalignments can occur due to an unstable measurement setup or a jitter-based countermeasure (e.g. by creating an unstable clock signal \cite{Guneysu_11}). Since \acp{CNN} are, to some extent, robust against variations in the input space as explained in Section \ref{sec:cnn}, our goal is to evaluate the image-based \ac{SCA} approach under such conditions.
We cut out segments of 180 sample points length from the traces for creating the images in order to ensure that enough leakage information is available.

\subsubsection{Hardware AES with high noise}
Another range of experiments are based on the public data set of the 
\textit{DPA Contest v2} \cite{dpav2}. These side-channel traces were acquired from an unprotected AES design running on an FPGA platform. The used AES module performs one round per clock cycle. Each trace contains 3253 sample points and covers a complete encryption operation. We have pre-selected 500 sample points for the experiments in order to attack the last round of the encryption. Several studies noted that this data set is relatively noisy and hard to attack \cite{picek_18}.

\subsubsection{Protected Software AES}
The last data set we use for the evaluation has been obtained from the public \ac{SCA} database \textit{ASCAD} \cite{Prouff_18}. The targeted platform is a first-order secured implementation of the \ac{AES} running on an 8-bit ATMega8518 microcontroller. Each trace is composed of 700 sample points and the targeted intermediate result is the third byte of the masked S-Box output that is processed during the first round. However, for training of the \ac{CNN}, we only use the unmasked result of the S-Box as labels to check whether the network is able to learn an higher-order attack using the information encoded in the images. For that purpose, the \ac{CNN} has to combine the leakages of the masks with the leakages of the masked S-Box. Considering the \ac{SNR} analysis given in \cite{Prouff_18}, those leakages occur roughly between the sample points $140-240$ and $450-550$. We therefore only used the respective parts of the traces as input for the 2D transformations and removed the remaining data points.

\subsection{Methodology}
After the traces have been encoded as images according to Section \ref{sec:image}, the \ac{CNN} described in Table \ref{tab:cnn} has been trained to solve the multi-class classification problem defined in Section \ref{sec:sca}.
As evaluation metric, we have used the well-known \ac{KGE} or key rank function. 
The \ac{KGE} quantifies the difficulty to retrieve the correct value of the key byte regarding the required number of attack traces \cite{Standaert_09}. It is calculated by replacing the probabilities on the right-hand side of equation (\ref{eq:max_lh}) by its approximations retrieved from the \ac{CNN}, and a ranking of the score vector after the evaluation of each attack trace. Intuitively, the faster the key rank converges to one, the more powerful is the attack.
In order to reduce the statistical bias that is induced by the random initialization of the \ac{CNN} weights, we have trained the network from scratch for ten times. During each run, we performed five independent attacks using 2000 images/traces from  $\mathcal{D}_{Attack}$. Finally, the mean key rank over all ten runs has been calculated which is similar to a ten-fold-cross validation evaluation as defined in \cite{Prouff_18}.

We have implemented our attack framework in Python using the open-source \ac{DL} framework \textit{Keras} \cite{_keras}. 
Training of the \ac{CNN} has been performed on a single Nvidia Tesla V100 GPU. As a baseline, we also executed the attack using the raw trace segments by replacing the 2D convolution and pooling operations in our \ac{CNN} with their 1D equivalents. Additionally, we compared our results to three different types of (unsupervised) 1D preprocessing techniques which are widely used in side-channel analysis for comparison: \ac{SSA}, \ac{FFT}, and \ac{PCA} using the 15 highest ranked components (this choice of parameter gave the best results in our experiments).

\subsection{Results for Single Image Attacks}
\label{sec:single_attack}

\begin{figure*}[h!]
	\centering
	\captionsetup[subfigure]{labelformat=empty}
	\subfloat[]{\scalebox{.5}{\includegraphics[bb=0 0 250 150]{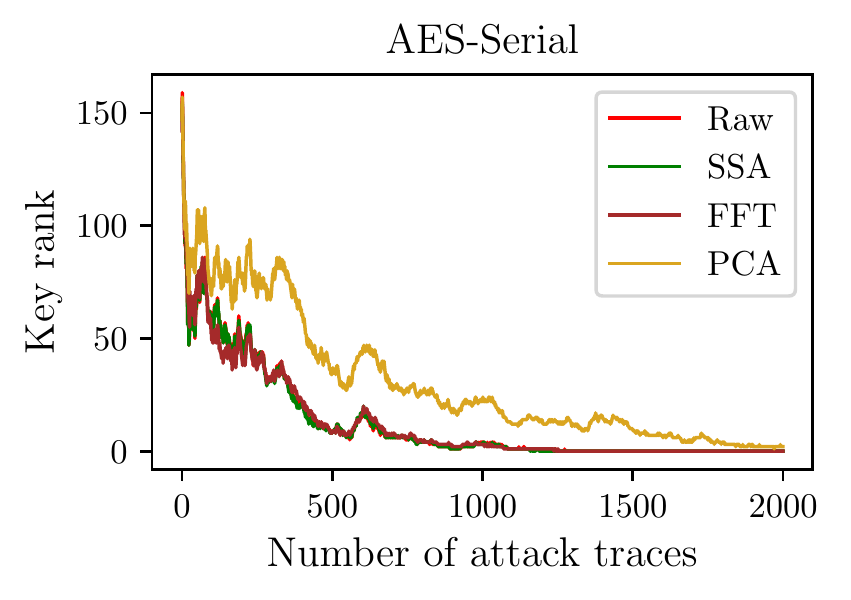}}\label{fig:kge_serial_1d}}
	\subfloat[]{\scalebox{.5}{\includegraphics[bb=0 0 250 150]{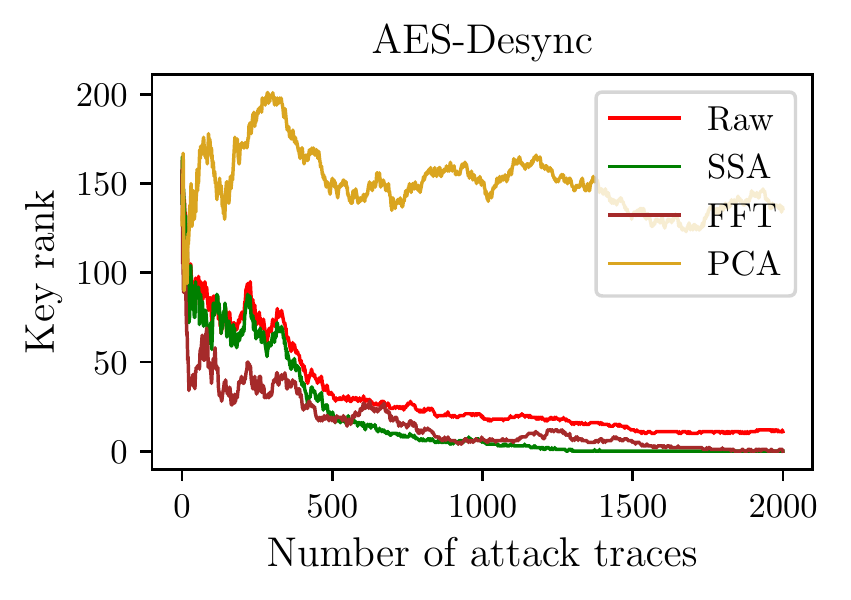}}\label{fig:kge_desync_1d}}
	\subfloat[]{\scalebox{.5}{\includegraphics[bb=0 0 250 150]{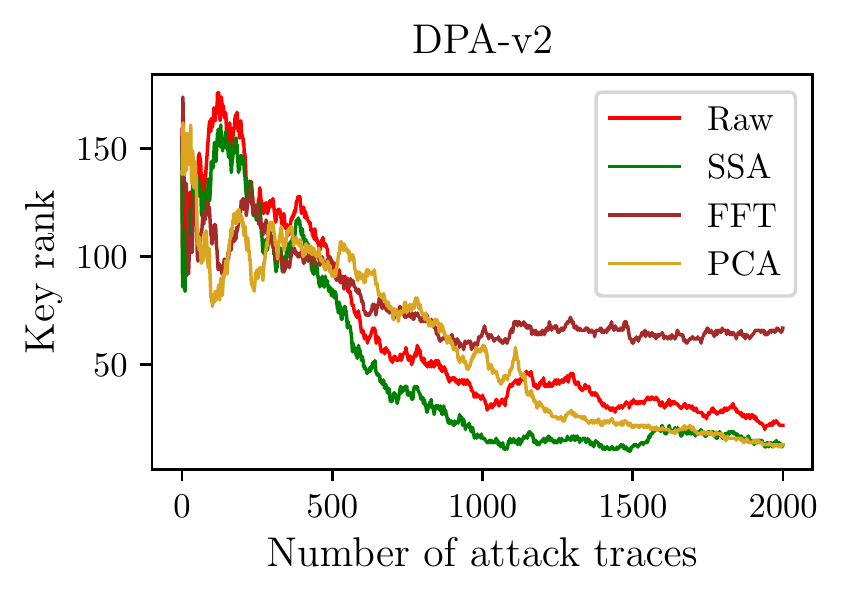}}\label{fig:kge_dpa_1d}}
	\subfloat[]{\scalebox{.5}{\includegraphics[bb=0 0 250 150]{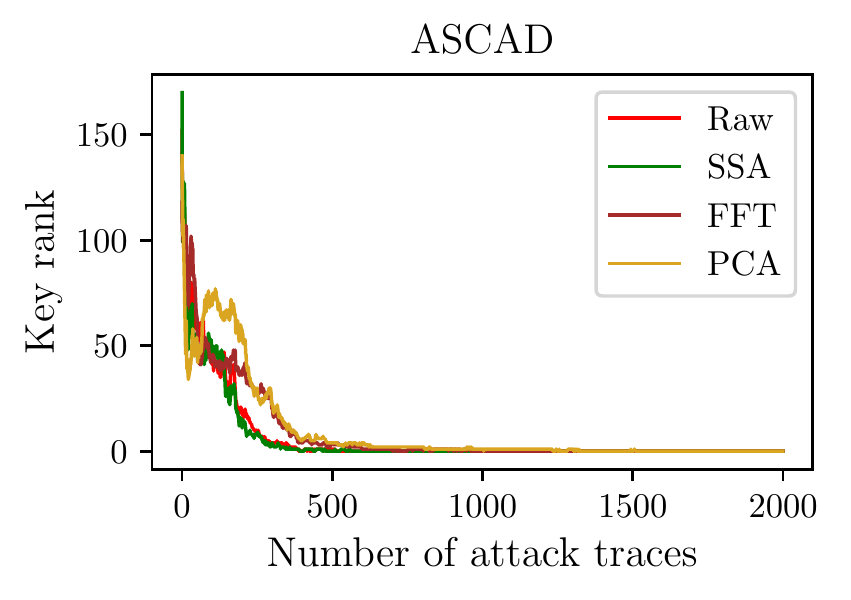}}\label{fig:kge_ascad_1d}}
	\hfil
	\subfloat[]{\scalebox{.5}{\includegraphics[bb=0 0 250 150]{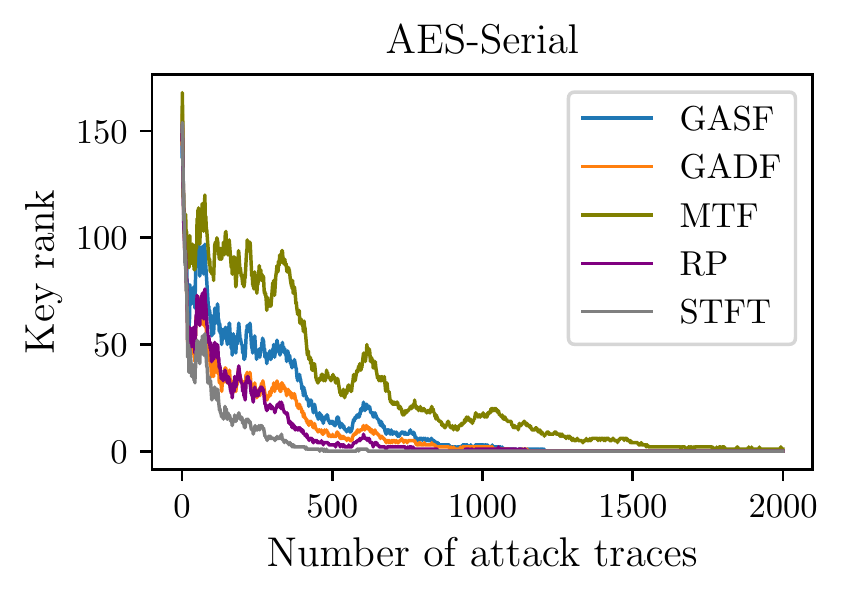}}\label{fig:kge_serial_2d}}
	\subfloat[]{\scalebox{.5}{\includegraphics[bb=0 0 250 150]{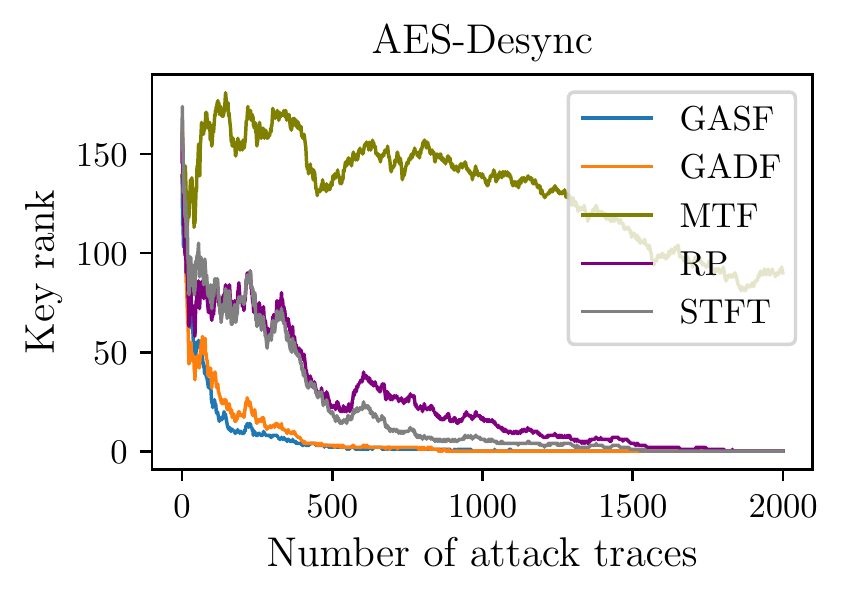}}\label{fig:kge_desync_2d}}
	\subfloat[]{\scalebox{.5}{\includegraphics[bb=0 0 250 150]{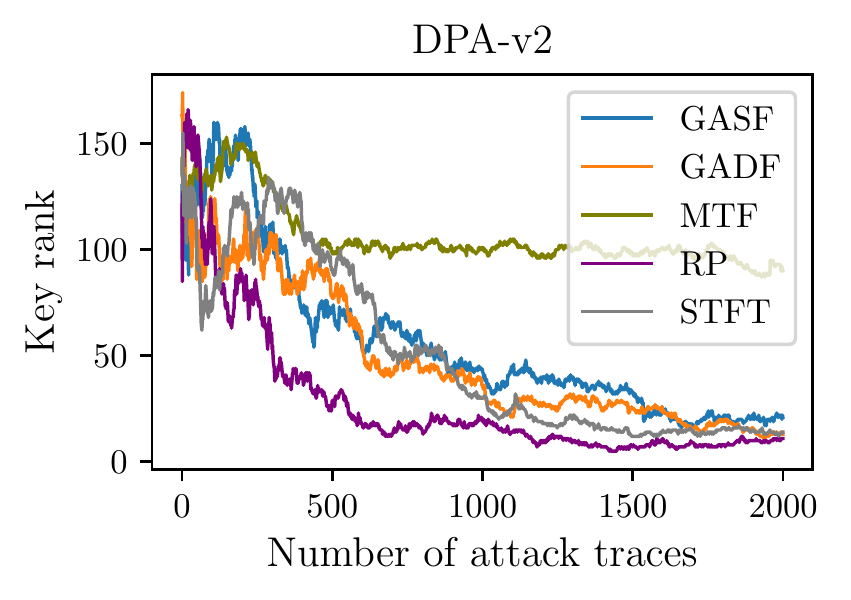}}\label{fig:kge_dpa_2d}}
	\subfloat[]{\scalebox{.5}{\includegraphics[bb=0 0 250 150]{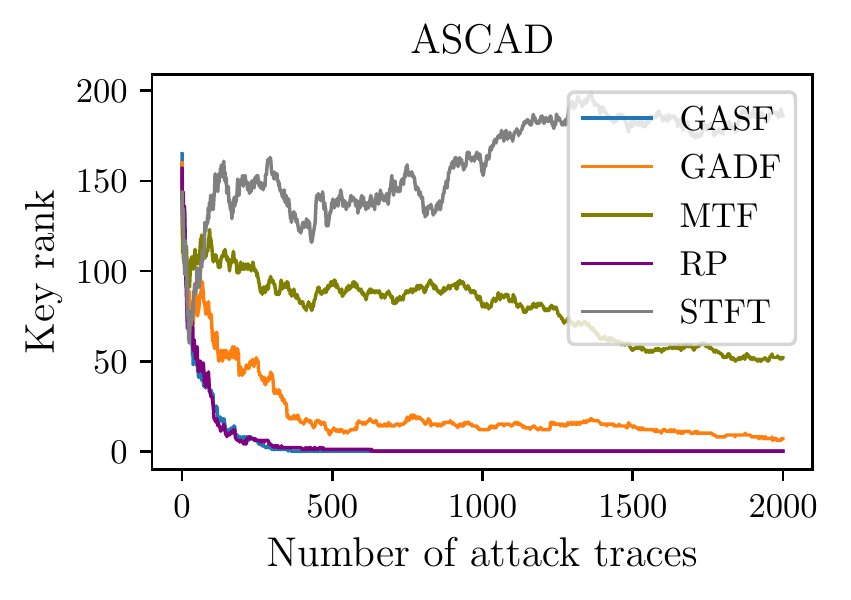}}\label{fig:kge_ascad_2d}}
	\hfil
	\subfloat[]{\scalebox{.5}{\includegraphics[bb=0 0 250 150]{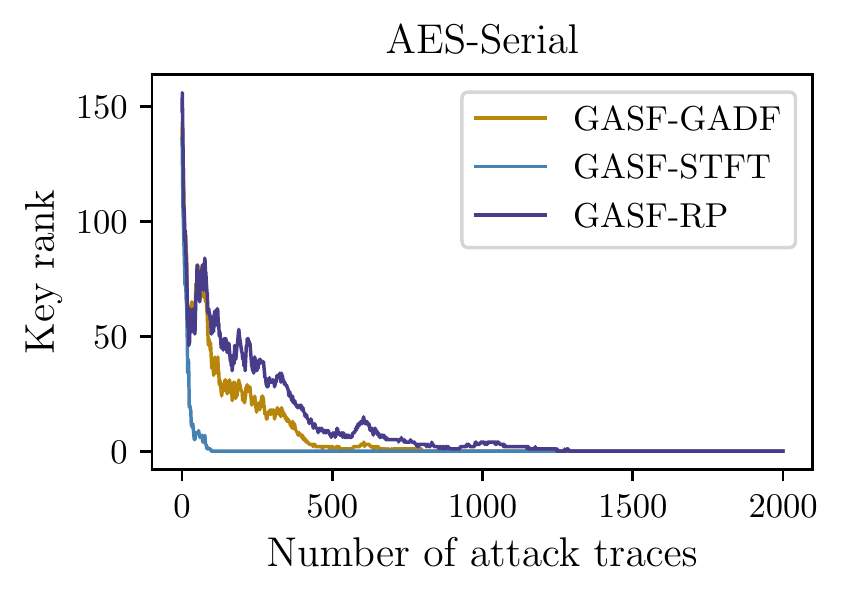}}\label{fig:kge_serial_2d_comb}}
	\subfloat[]{\scalebox{.5}{\includegraphics[bb=0 0 250 150]{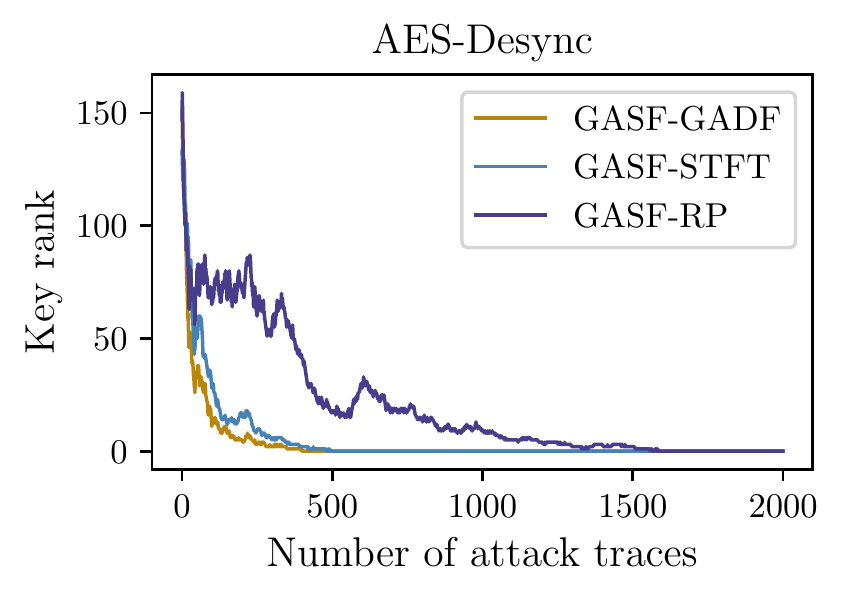}}\label{fig:kge_desync_2d_comb}}
	\subfloat[]{\scalebox{.5}{\includegraphics[bb=0 0 250 150]{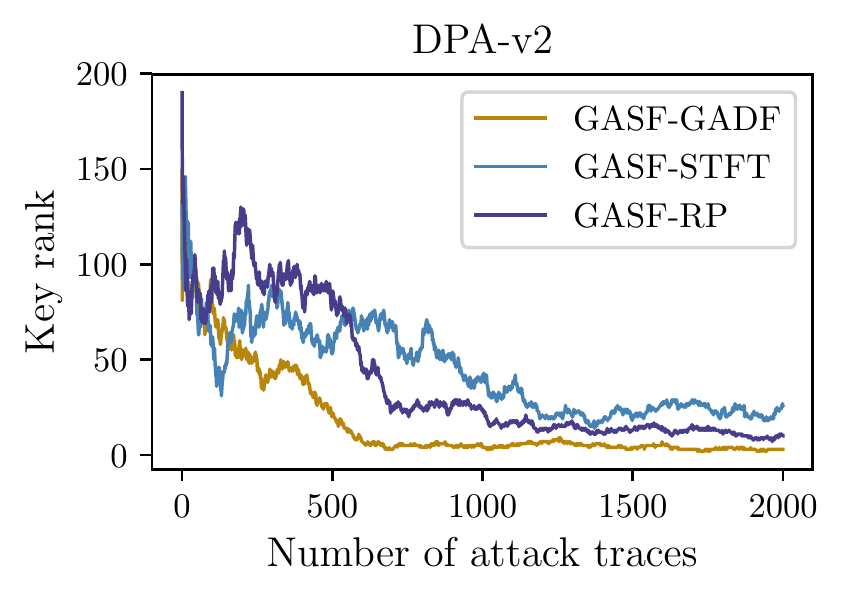}}\label{fig:kge_dpa_2d_comb}}
	\subfloat[]{\scalebox{.5}{\includegraphics[bb=0 0 250 150]{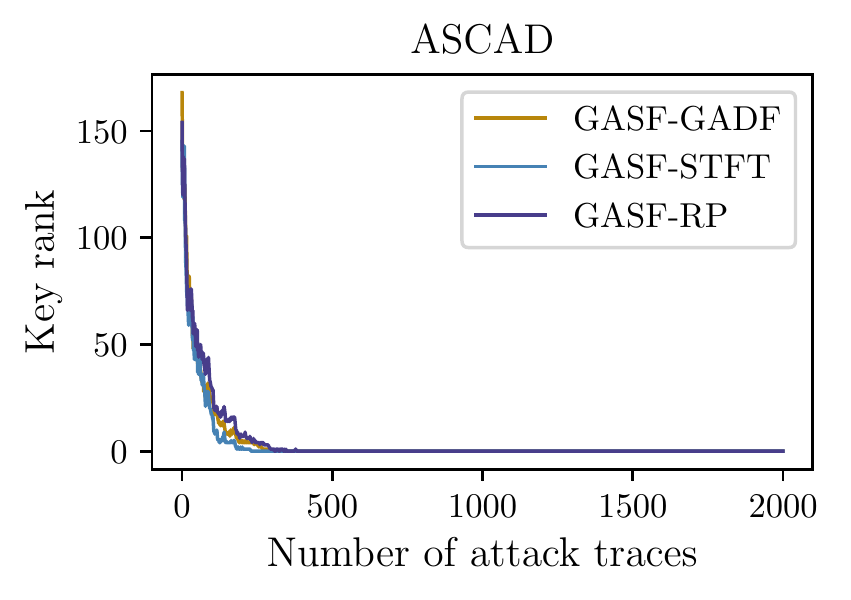}}\label{fig:kge_ascad_2d_comb}}
	\caption{Overview of attack results over the four data sets. The mean \ac{KGE} for the 1D baseline (raw traces) along with the 1D transformation methods are illustrated in the first row. The second row shows the mean key ranks for the single image attacks, while the combined image attack results are illustrated in the third row.}
	\label{fig:kge}
\end{figure*}

\figurename{~\ref{fig:kge}} shows the results for the 1D baseline attacks in the first row, and the single image-based attacks in the second row. From there, one can observe the following:

\subsubsection{\ac{GAF}}
The \ac{GAF} methods show almost similar results in the AES-Serial, AES-Desync and DPA-v2 data sets, while \ac{GASF} shows a considerably steeper key rank curve than \ac{GADF} for the protected software AES. We expected a larger difference between the two \ac{GAF} variants since \ac{GASF} images encode information about the original power trace in its main diagonal which is not directly present in \ac{GADF} images. It seems that the temporal correlations contained in \ac{GAF} images alone provide enough features to recover the secret key. 
Another notable point is the good performance on the desynchronized data set. While the remaining transformation techniques and also the 1D attacks need more traces to reach a key rank of 1 compared to the aligned traces, \ac{GASF} and \ac{GADF} show a better \ac{KGE}.
This can be explained by the representation of the 1D signal in polar coordinates, which helps to preserve absolute temporal relations. Additionally, the spatial diversity of features in \ac{GAF} images provide a regularization factor that is known from data augmentation methods. This improves the \ac{CNN} model's ability to generalize distortions in the test set. We will exploit this aspect later in the paper to further enhance the performance of \ac{GAF}-based attacks.

\subsubsection{\ac{RP}}
The {RP} transform performs best in two out of the four data sets among the image-based attacks, and is able to consistently beat the 1D baseline. Except for the AES-Desync data set, it furthermore performs similar or better than the best 1D preprocessing method \ac{SSA}. 
This proves that recurrence as defined in equation (\ref{eq:rp}) is a suitable method to represent variations in the power consumption, which can be exploited effectively by a 2D \ac{CNN} to distinguish between different values of the attacked operation.

\subsubsection{\ac{MTF}}
Looking at the result for \ac{MTF}, one can see that it has a significantly worse key rank than the other approaches in all four data sets. One reason is the reduction of temporal dependencies through the quantization step, which is required for building the transition matrix. Another reason is that the \ac{MTF} transform produces a non-bijective mapping, i.e, reconstructing the raw trace from an \ac{MTF} image in an accurate manner is - in contrast to \ac{GAF} - hardly possible. However, previous work showed that this property is an important factor for classifying time series images. The observation that \ac{GAF} generally performs superior to \ac{MTF} is thus in line with other published results \cite{Wang_15} and our correlation analysis shown before.

\subsubsection{\ac{STFT}}
The best performance on the AES-Serial data set is achieved by the \ac{STFT}-based attack. Although the spectrograms only contain five frequency bands, the additional information helps to reduce the amount of required attack traces from 1000 to 500 in order to achieve a KGE of one compared to the 1D baseline attacks. An improved \ac{KGE} can also be reported for the desynchronized AES traces and the noisy DPA-v2 data set.
However, the spectrogram-\ac{CNN} fails to attack the first-order secured implementation. We initially assumed that a window length of eight in \ac{STFT} may result in a too small frequency resolution for the ASCAD data set, as the traces have been sampled with a rate of 2 GS/s. However, we got almost the same result when using larger \ac{STFT} window sizes (16, 64, 96, 128) following the guidelines of Yang et al. \cite{Yang_18}. We will later see that not the \ac{STFT} parameters are the reason for the bad performance, but the way how the spectrogram is given to the \ac{CNN}.

\subsection{Results for Combined Image Attacks}
The transformation techniques presented above encode different types of information about time series data (like power measurements): \ac{GASF}/\ac{GADF} show the superposition/difference of directions at two time steps, \ac{MTF} images encode transition probabilities from quantile to quantile, \acp{RP} reveal space dynamics as distance between corresponding trajectories, and spectrograms visualize the energy of a signal as a function of frequency and time. Thus, we can consider them as complementary representations of the same data source which can be combined into a single image to improve the classification performance of the \ac{CNN}. The combination has been done by concatenating images generated from the corresponding power trace along the z-axis as shown in the center of \figurename{~\ref{fig:overview}}. Except for spectrograms, this was easily possible since the images are of identical vertical and horizontal resolution. To join spectrograms with other images, we have upscaled the y-axis by dividing the five frequency bands until the proper resolution has been reached (i.e. a rectangular shape).

We have concentrated our evaluation on \ac{GASF}-\textit{X} combinations due to the good performance of \ac{GASF} on single image attacks. The results are illustrated in the third row of \figurename{~\ref{fig:kge}}. From there, one can notice the following:

\subsubsection{GASF-RP}
This combination yields a small enhancement of the \ac{KGE} in the attack against the protected software \ac{AES}, likely because both techniques individually perform well on this data set. The results on the other data sets are comparable or even worse than single image \ac{RP} attacks. This indicates that \ac{RP} and \ac{GASF} features are too diverse for the early data fusion approach that we have applied in this work. The analysis of a combination of \ac{RP} and \ac{GASF} features in a deeper \ac{CNN} layer (i.e. late information fusion \cite{Eitel_15}) is subject to future work.

\subsubsection{GASF-STFT}
The setup of rescaled spectrograms and \ac{GASF} images shows the best \ac{KGE} in two out of four data sets. The margin to the remaining methods is especially large in the AES-Serial data, where a stable key rank of one is reached with less than 100 attempts. This means an improvement of a factor of ten compared to the 1D baseline and single \ac{GASF}, and a factor of five compared to unscaled spectrograms. 
The performance increase for the ASCAD data set is is also remarkable considering the worse result of the spectrogram-only attack.
We found that this mismatch is mainly due to the frequency upscaling of spectrograms needed for the image fusion.
It improves the \ac{CNN} kernel's ability to capture transitions between different frequency bands. While in the unscaled spectrograms all frequency bands have been processed together by the kernel in the first layer, in the upscaled variant at most two frequency bands are evaluated in parallel. This enables the kernels to create more discriminant and informative features for deeper layers. Indeed we could examine that the scaled spectrograms alone give a significantly improved \ac{KGE} compared to the unscaled version, and also with regard to spectrograms having a higher frequency resolution that were generated with a larger \ac{STFT} window parameter.
However, similar to GASF-RP images, there seems to be no improvement on the DPA-v2 data set. These findings suggest that a concatenation of different image types is of less benefit in case of highly noisy setups.

\subsubsection{GASF-GADF}
The combination of the \ac{GAF} images shows remarkable results for the AES-Desync and DPA-v2 data sets. This makes it very suitable for attacks on misaligned and noisy traces. The combination is the only one which could improve the \ac{KGE} over all data sets compared to single image attacks. furthermore, GASF-GADF is the only combination that is able to perform better than the 1D-based attacks in all setups.
The \ac{GAF} transformations have different representations of temporal relations which rely on polar coordinates. Therefore, it seems that the CNN is able to detect features in the combined representations of temporal correlations to enhance the classification accuracy for all attacked data sets.

In order to compare the results of the combined GASF-GADF attack with other work, we show the \ac{KGE} of related approaches in Table \ref{tab:comparison}. We focus the comparison on the ASCAD database since it represents a de-facto benchmark for \ac{DL}-based \acp{SCA}. Although the approaches are not completely comparable due to changing evaluation setups, one can still notice that the image combination attack requires significantly less traces to approach a stable key rank smaller than two. This further motivates the applicability of our approach despite the timing overhead that is introduced through the transformation of the traces into images (i.e. training time per epoch is roughly doubled on our setup compared to the 1D attacks against ASCAD).

\begin{table}
	\centering
	\caption{Comparison of attacks for the ASCAD data set}
	\label{tab:comparison}
	\begin{tabular}{ccc}
		\toprule
		\textbf{Source} & \textbf{\ac{KGE} $\leq 2$} & \textbf{Evaluation Method}\\
		\midrule
		\cite{Prouff_18} & $\approx 450$ & 10-fold-cross validation \\
		\cite{Kim_2019} & $\approx 500$ & Repeated random subsampling \\
		\cite{Robyns_2018} & $\approx 700$ & 10-fold-cross validation \\
		\cite{Pfeifer_2018} & 338 & Mean rank (mask values revealed) \\
		This Paper & $ \approx 275$ & 10-fold-cross validation \\

		\bottomrule
	\end{tabular}
\end{table}

\subsection{Image-based Data Augmentation}
\label{sec:augmentation}

\begin{figure*}[t]
	\centering
	\subfloat[]{\scalebox{.5}{\includegraphics[bb=0 0 250 150]{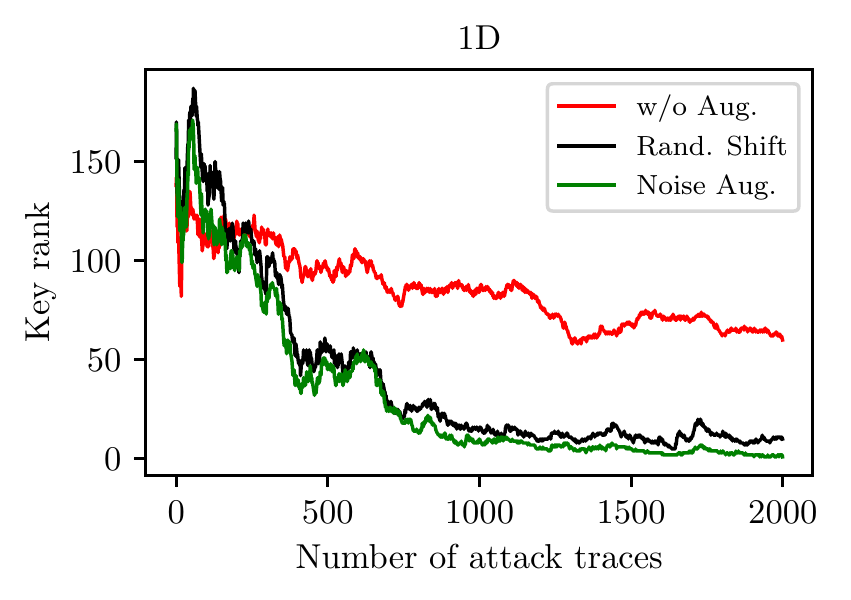}}\label{fig:aug_1d}}
	\subfloat[]{\scalebox{.5}{\includegraphics[bb=0 0 250 150]{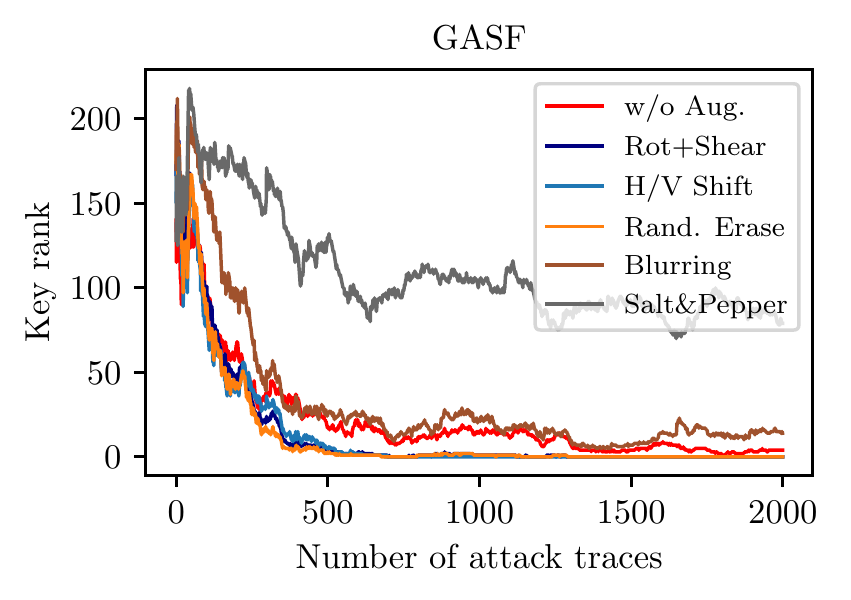}}\label{fig:aug_gasf}}
	\subfloat[]{\scalebox{.5}{\includegraphics[bb=0 0 250 150]{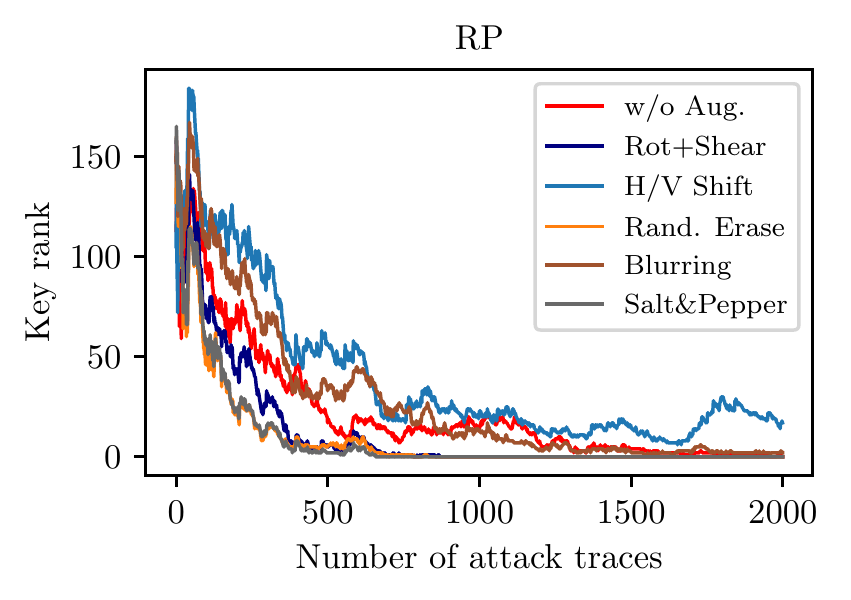}}\label{fig:aug_rp}}
	\subfloat[]{\scalebox{.5}{\includegraphics[bb=0 0 250 150]{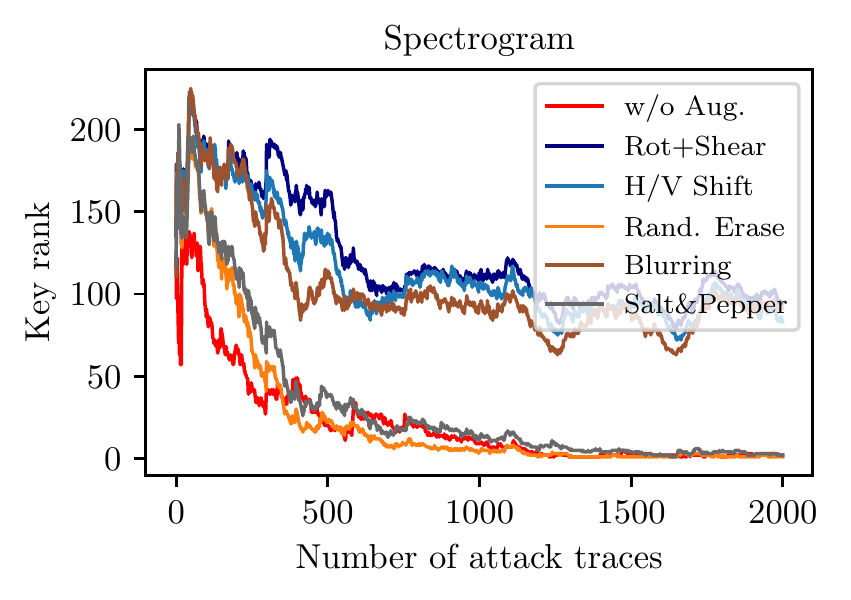}}\label{fig:aug_stft}}
	\caption{Comparison of trace- and image-based data augmentation methods using $S_P = 1000$ training examples: (a) 1D Trace, (b) GASF, (c) RP, (d) Spectrogram. The red curves in the plots correspond to attacks without augmentation.}
	\label{fig:kge_aug}
\end{figure*}

Data augmentation can be seen as a way to artificially increase the training set by showing the model many variations of the same input data during training. It is an extensively used technique in object classification due to the manifold possibilities to apply geometric modifications to images. The augmentation techniques do not change the label of an image, hence allow the model to better generalize \cite{Goodfellow_2016}.
This section aims to evaluate if 2D representations of power traces can also benefit from image-based data augmentation, especially in cases where the size of the profiling set is rather small. For that purpose, we have restricted the number of training data to $S_P = 1000$, and applied the following five image preprocessing techniques:
\begin{itemize}
	\item Random rotation by up to 40 degrees, plus a shearing operation with intensity 0.5
	\item Random horizontal and vertical shift by up to 20\% of the image size
	\item Erasing randomly picked rectangular patches (pixels are set to random values between 0 and 1)
	\item Gaussian blurring
	\item Random addition of salt and pepper noise (i.e. individual pixels are set to either zero or one)
\end{itemize}
The methods have been selected since they represent different classes of augmentation types (geometric and arithmetic operations, segmentation, blurring).
We have individually evaluated the effect of each augmentation technique on all kinds of image transformations except \ac{MTF} (due to the reasons discussed earlier in the paper). Additionally, we have implemented the random shift technique proposed in \cite{Cagli2017} and the noise addition method of Kim et al. \cite{Kim_2019} to compare our results with state-of-the-art 1D augmentation techniques. 
The results are shown in \figurename{~\ref{fig:kge_aug}}. From there, it can be observed that the augmentation methods perform differently on the imaging types. For example, the geometric operations (rotation, shifting) lead to good results when applied to \ac{GASF} images, but deteriorate the key rank of spectrogram-based attacks. This can be explained by the usage of polar coordinates in \ac{GAF} images. Our spectrograms have a very low frequency resolution, thus shifting or rotating introduces too much variance in the training set.
Random erasing, on the other hand, gives a notable performance boost in all attacks and seems to be a suitable standard choice. It introduces an explicit form of regularization by adding block noise, but maintains the global structure of the image. This feature already showed good results on various recognition tasks \cite{Zhong_17}. 

When comparing image-based augmentation with trace-based augmentation, one can easily notice that image-based augmentation is superior. While the 1D attacks only manage a key rank between 15 and 20 using 2000 encryptions, some of the \ac{GASF} and \ac{RP}-based attacks have a stable \ac{KGE} below two with only 700 traces. This means, by using image augmentation, we achieved better results in the \ac{GASF} and \ac{RP}-based attacks than in the experiments of Section \ref{sec:single_attack} with only $1/50$ of the training set. Another interesting point to notice is the large gap between the curves without augmentation. The imaging methods are still able to converge to a rank below five, while the 1D attack reaches a key rank of around 60 after 2000 attempts. This is an unexpected result as the 2D CNN has ten times the amount of trainable parameters compared to the 1D version. Thus, the 2D model is much more prone to overfitting in our restricted setting with only 1000 training samples. However, the 2D attacks still outperform the 1D attacks (even with trace augmentation). Overall, the results of this section clearly confirm the effectiveness of \ac{DL}-based \acp{SCA} in two-dimensional space and the increased information content of the generated images.

\section{Conclusion}
In this work, we have examined several transformations of 1D power traces into a 2D space in order to improve the efficiency of \ac{DL}-based \acp{SCA}. The performance has been evaluated against advanced 1D representation techniques using four different data sets. Among the considered methods, \ac{GAF} and \ac{RP} have shown the best results in single image attacks. Furthermore, we have compared different combinations of images, which resulted in a further enhancement of the classification. The combination of  \ac{GASF}-\ac{STFT} has shown a ten-fold improvement on the AES-Serial implementation over the 1D baseline, and \ac{GASF}-\ac{GADF} has proven remarkable results on both desynchronized and noisy AES traces. These two combinations have demonstrated very robust results on all data sets and outperform related approaches on the public ASCAD database. Finally, we have investigated the impact of data augmentation on the transformed traces, and were able to observe the superiority of imaging methods over state-of-the-art 1D attacks.

As a future work, we intend to evaluate other types of data augmentation techniques, as well as a combination of different techniques. Another interesting path could be to explore the effect of image feature fusion in a deeper layer of the \ac{DNN}, instead of the image stacking approach we have applied in this work. We expect these methods to further improve the promising results of this paper.
\section{ACKNOWLEDGMENT}
{This work is supported in parts by the German Federal Ministry of Education and Research (BMBF) under grant agreement number 16KIS0606K (Security by Reconfiguration - SecRec)

\bibliographystyle{IEEEtran}
\bibliography{IEEEabrv,main}
\end{document}